# Ultrathin, high-speed, all-optical photoacoustic endomicroscopy probe for guiding minimally invasive surgery

TIANRUI ZHAO,[1] TRUC THUY PHAM[1] CHRISTIAN BAKER,[1] MICHELLE T. MA,[1] SEBASTIEN OURSELIN,[1] TOM VERCAUTEREN,[1] EDWARD ZHANG,[2,3] PAUL C. BEARD,[2,3] AND WENFENG XIA[1,*]

[1]*School of Biomedical Engineering and Imaging Sciences, King's College London, 4th Floor, Lambeth Wing St Thomas' Hospital, London SE1 7EH, United Kingdom*
[2]*Department of Medical Physics and Biomedical Engineering, University College London, Gower Street, London WC1E 6BT, UK*
[3]*Wellcome/EPSRC Centre for Interventional and Surgical Sciences, University College London, Charles Bell House, 67-73 Riding House Street, London W1W 7EJ, UK*
[*]*wenfeng.xia@kcl.ac.uk*

**Abstract:** Photoacoustic (PA) endoscopy has shown significant potential for clinical diagnosis and surgical guidance. Multimode fibres (MMFs) are becoming increasing attractive for the development of miniature endoscopy probes owing to ultrathin size, low cost and diffraction-limited spatial resolution enabled by wavefront shaping. However, current MMF-based PA endomicroscopy probes are either limited by a bulky ultrasound detector or a low imaging speed which hindered their usability. In this work, we report the development of a highly miniaturised and high-speed PA endomicroscopy probe that is integrated within the cannula of a 20 gauge medical needle. This probe comprises a MMF for delivering the PA excitation light and a single-mode optical fibre with a plano-concave microresonator for ultrasound detection. Wavefront shaping with a digital micromirror device enabled rapid raster-scanning of a focused light spot at the distal end of the MMF for tissue interrogation. High-resolution PA imaging of mouse red blood cells covering an area 100 $\mu$m in diameter was achieved with the needle probe at ~3 frames per second. Mosaicing imaging was performed after fibre characterisation by translating the needle probe to enlarge the field-of-view in real-time. The developed ultrathin PA endomicroscopy probe is promising for guiding minimally invasive surgery by providing functional, molecular and microstructural information of tissue in real-time.

## 1. Introduction

Optical endoscopy is commonly used for the diagnosis of diseases inside human body. However, conventional endoscopes using white-light-illumination only provide morphological information of superficial tissues and thus, biopsy is usually needed to extract small pieces of tissues from suspicious regions for definitive histopathological analysis. This process is time-consuming, increases costs, and may remove functional tissue and miss abnormalities due to sampling errors. Various advanced endoscopy modalities have been investigated to provide tissue characterisation *in situ* as termed 'optical biopsy' [1]. Endoscopic optical coherence tomography (OCT) enables real-time three dimensional (3D) imaging of tissues with microscopic scale (cellular) morphological information derived from optical scattering of biological tissues, however, it is challenging to distinguish between different tissue types due to the lack of molecular contrast [2]. In contrast, fluorescence imaging employs specific fluorescent labels to highlight abnormal tissues such as tumour cells, but it is difficult to perform in-depth imaging [3]. As a hybrid imaging modality, photoacoustic (PA) imaging inherits advantages from both optical and ultrasound imaging, providing both depth-resolved structural and molecular information of tissue by optically exciting ultrasound waves from tissue chromophores [4–7]. Furthermore, PA imaging with multispectral illumination can further provide accurate functional information such as blood oxygen saturation, which enables imaging of tumour hypoxia and metabolism *in vivo*. As such, photoacoustic endoscopy (PAE) has attracted significant interest for *in situ* diagnosis of tissue. In the last decade, various photoacoustic endoscopy (PAE) probes have been developed for intravascular and gastrointestinal tract imaging [7–11]. Although most probes have previously been side-viewing, a number of forward-viewing probes have been developed recently for several minimally invasive procedures including tumour biopsy and fetal interventions where tissue characterisation in front of the surgical device is required [12, 13]. Early forward-viewing PAE probes used coherent fibre bundles with raster-scanning of a focused laser beam at its distal end to perform optical-resolution photoacoustic microscopy (OR-PAM) imaging of tissue [14–16]. The lateral resolution was limited to ∼ 7$\mu$m due to the gaps between individual fibre cores. In recent years, Ansari *et al.* developed several forward-viewing PAE probes based on the PA tomography principle. In a recent study [12], a bichromatic Fabry-Perot sensor was coated at the tip of a rigid fibre bundle and raster-scanning of an interrogation laser beam through the bundle was performed for ultrasound detection. Most recently, a Fabry-Perot sensor was interrogated through a flexible fibre bundle with a miniature optical relay system [13]. Vasculatures of duck embryo and human placenta *ex vivo* were visualised down to a depth larger than 1 mm with a spatial resolution of several tens of micrometres. The PA tomography approach provided a greater tissue penetration depth compared to those with the OR-PAM counterpart, but at the expense of spatial resolution.

Recently, in order to scan a focused spot over a target located at the distal end of a fibre, MMFs were studied as an alternative to fibre bundles by focusing laser light through the fibres via wavefront shaping [17, 18]. Light transmission characteristics through a MMF were measured with a spatial light modulator, which can be subsequently used to shape the output light into a tightly focused beam by modulating the input light field and raster-scan it at the distal MMF tip for imaging [17, 18]. Compared to fibre bundles, MMF-based PAE benefits from greater pixel densities, thinner probe sizes, and lower costs. More importantly, the MMF-based endoscope provides a greater flexibility as the focal spot diameter, shape and focal plane can be adjusted as compared to bundle-based OR-PAM implementations. In 2020, Mezil et al. fabricated a dual-modal PA/fluorescence probe based on a MMF with a fibre-optic ultrasound sensor for ultrasound detection [19]. However, it took 30 s for single frame acquisition due to the use of a slow liquid-crystal spatial light modulator (LC-SLM) for wavefront shapinng. Digital micromirrorr devices (DMDs) have been studied as a fast alternative to LC-SLMs to improve the speed of wavefront shaping [20–22]. In a recent study, we demonstrated high-speed PA/fluorescence imaging through a MMF fibre via wavefront shaping using a fast DMD [23]. However, the system used a bulky piezoelectric transducer for ultrasound detection in a transmission mode, where tissue samples were placed between the MMF tip and the transducer, and thus limited its usability.

In this work, we developed a miniature forward-viewing PA endomicroscopy probe integrated within the cannula of a medical needle for guiding minimally invasive procedures. This needle probe comprises a MMF for PA excitation laser delivery and a highly-sensitive fibre-optic microresonator sensor for ultrasound detection. As wavefront shaping employs a high-speed DMD, the imaging speed was improved by more than two orders of magnitude as compared to that with a LC-SLM.

## 2. Materials and Methods

### 2.1. Photoacoustic endomicroscopy setup

A schematic diagram of the imaging system is shown in Fig. 1 (a). A pulsed laser emitting at 532 nm (2 ns, SPOT-10-200-532, Elforlight, UK) was used as the PA excitation light source. A 30 cm-long gradient index (GRIN) fibre ($\phi$100 $\mu$m, 0.29 NA, Newport, California) was used for the delivery of the excitation light. A DMD (768 × 1080 pixels, DLP7000, Texas Instruments, Texas) was used to project binary patterns onto the proximal end of a MMF via an achromatic doublet lens (f = 50 mm, AC254-050-A-ML, Thorlabs, New Jersey) and an objective (20×, 0.4 NA, RMS20×, Thorlabs, New Jersey). A sub-region of the DMD covering 128×128 micromirrors was employed.

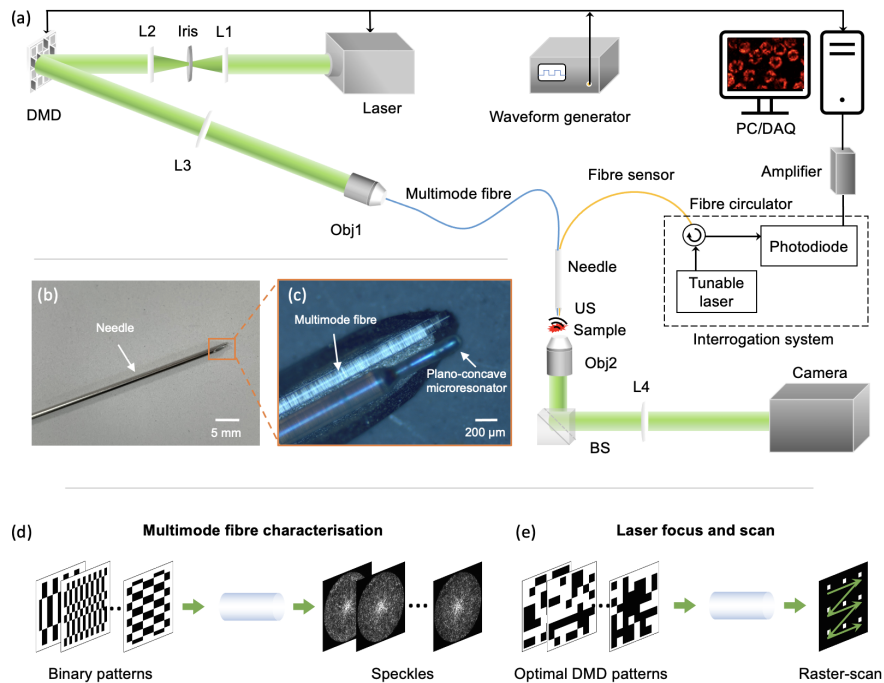

Fig. 1. Illustration of the photoacoustic endomicroscopy imaging system. (a)Schematic diagram of the experimental setup. L1-4, achromatic doublet lenses; DMD: digital micromirror device; US, ultrasound; Obj1-2: Objective lenses; BS: beamsplitters; PC: personal computer; DAQ: data acquisition. (b) A photo of the imaging probe integrated within a spinal needle (20 gauge). The needle has an outer diameter of 0.9 mm and an inner diameter of 0.6 mm. (c) A microscopy image of the needle tip region. Schematic diagrams of the principles of (d) multimode fibre characterisation and (e) raster-scanning of a focused laser beam through the multimode fibre.

PA signals (ultrasound waves) were detected by a fibre-optic ultrasound sensor based on a plano-concave microresonator at the tip of a single-mode fibre. The microresonator comprised a dome-shaped epoxy spacer sandwiched by two dichroic mirrors, which was interrogated by a wavelength-tuneable continuous wave laser (TSL-550, Santec, UK). The incident ultrasound waves deform the epoxy spacer leading to changes in the optical reflectivity of the microresonator [24, 25]. An optical circulator (6015-3-APC, 1525-1610 nm, Thorlabs, New Jersey) was employed to deliver the interrogation laser to the microresonator cavity and collect the reflected light using a photodiode (G9801-22, Hamamatsu, Shizuoka Pref. Japan). The output was connected to a data acquisition card (M4i.4420, Spectrum Instrumentation, Grosshansdorf, Germany) after amplification (SPA.1411, Sprectrum Instrumentation, Grosshansdorf, Germany) and then transferred to a personal computer (Intel i7, 3.2 GHz) for processing. Compared to conventional piezoelectric ultrasound transducers, the optical sensor was nearly omni-directional and had a large bandwidth and a high sensitivity with a small size (125 $\mu$m in diameter) [24, 25].

Synchronisation of the DMD patterns display, laser-firing, and data acquisition was controlled by a waveform generator (33600A, Keysight, Santa Rosa, California) and a custom MATLAB program. Both the MMF and the fibre-optic ultrasound sensor were integrated within the cannula of a spinal needle (20 gauge), which was affixed on a 3D translation stage. The needle tip (Fig. 1b and c) was inserted into a custom imaging tank filled with deionised water for acoustic coupling.

*2.2. Multimode fibre characterisation*

Since laser transmission through a MMF is scrambled due to modal dispersion [26], a characterisation process based on a real-valued transmission matrix (RVITM) was implemented prior to imaging. The characterisation process was reported in our previous study [27], in brief, a series of binary patterns were displayed with the DMD whilst the speckle patterns transmitted from the distal MMF tip were captured by a CMOS camera (C11440-22CU01, Hamamatsu, Shizuoka Pref. Japan) after magnification by an objective (20×, 0.4 NA, RMS20×, Thorlabs, New Jersey) and an achromatic doublet lens (f = 100 mm, AC254-0100-A-ML, Thorlabs, New Jersey). To create the binary patterns, a Hadamard matrix H∈(-1, +1) with dimensions of N×N was first generated in MATLAB, then two binary matrices $H_1 = (H + 1)/2$ and $H_2 = (-H + 1)/2$ were constructed. Each column of $[H_1, H_2]$ was then converted into a binary pattern to be displayed on DMD. The input and output light intensities through a MMF were modelled with a RVITM as:

$$\begin{bmatrix} I_1^1 & \cdots & I_1^{2N} \\ \vdots & \ddots & \vdots \\ I_m^1 & \cdots & I_m^{2N} \end{bmatrix} = RVITM \cdot [H_1, H_2], \tag{1}$$

where $I_m^k$ is the intensity at the $m^{th}$ output mode when the $k^{th}$ binary pattern is displayed as input, $N$ is the total number of input modes, and · represents matrix multiplication. To obtain the value of the RVITM, Eq. 1 was further expressed to arrive at a binary matrix $[H, -H] = [2H_1 - 1, 2H_2 - 1]$ on the right-hand side, and owing to the properties of a Hadamard matrix, $[H, -H]^T = [H, -H]^{-1}$, the value of RVITM can be calculated from intensity-only speckles and pre-known binary patterns via simple matrix manipulation as:

$$RVITM = \begin{bmatrix} 2I_1^1 - I_1^1 & \cdots & 2I_1^{2N} - I_1^1 \\ \vdots & \ddots & \vdots \\ 2I_m^1 - I_m^1 & \cdots & 2I_m^{2N} - I_m^1 \end{bmatrix} \cdot [H, -H]^T, \tag{2}$$

The expression of the transmission constant that connects the $n^{th}$ micromirror and the $m^{th}$ output mode can be further derived as $rvit_{mn} = A_{mn}A_R cos(\theta_{mn} - \phi_R)$, where $A_{mn}$ and $\theta_{mn}$ are the amplitude and phase at the $m^{th}$ output mode when only the $n^{th}$ micromirror is switched 'ON', $A_R$ and $\phi_R$ are the amplitude and phase of the output light field when all micromirrors are switched 'ON', respectively. It means that switching 'ON' micromirrors with positive $rvit_{mn}$ values leads to constructive interference because the phases of output fields are in the range of $[-\pi/2, \pi/2]$. To further increase the peak-to-background ratio, we ranked all the micromirrors descendingly with their $rvit_{mn}$ values and switched 'ON' the top 30% micromirrors as an optimal pattern for focusing at desired spatial locations at the distal end of the MMF as demonstrated in our previous study [28]. Then, by sequentially displaying the optimal patterns, a tightly focused laser beam can be raster-scanned in front of the distal tip of the MMF (Fig. 1 e). Optical sectioning was achieved by setting the focal plane of the camera to a series of desired optical focal planes during the fibre characterisation process. For each focal plane, a RVITM was calculated.

### 2.3. Imaging performance characterisation

After MMF characterisation, the imaging samples were placed in front of the MMF tip at the focal plane, and the excitation light was focused and raster-scanned over the samples by displaying optimal patterns on the DMD. PAM images were displayed as maximum intensity projections (MIPs) after signal denoising with a 20 MHz low-pass filter. Two metrics, the enhancement factor (EF) and the size of the focus, were used to characterise the focusing performance. The former largely determines the signal-to-noise ratio and the latter determines the lateral resolution of the imaging system. The EF was defined as the ration between the maximum light intensity in the focusing position and the average light intensity in the background. Accordingly, the power ratio was calculated as the fraction of total output light energy distributed in the focal area. The size of the focus was defined as the full-width-at-half-maximum (FWHM) value of the intensity profile across the centre of the focus.

The lateral resolution was measured by imaging a 1951 USAF Resolution Test Targets (Thorlabs, New Jersey, USA). PA images over a $\phi 100$ $\mu m$ field-of-view (FOV) were acquired with scanning step sizes of 0.5 and 1 $\mu$m, respectively. To estimate the lateral resolution, an edge spread function (ESF) was first obtained by averaging across profiles at 10 adjacent positions across an edge of a bar of the resolution target, and then a line spread function (LSF) was achieved by calculating the derivative of the ESF. The lateral resolution was calculated as the FWHM value of the Gaussian fit of the LSF. Since the point spread function suffers aberration at peripheral regions [29], the lateral resolution at both central and peripheral regions were measured.

Two methods of obtaining depth information in 3D-PA endomicroscopy imaging were compared: acoustic sectioning and optical sectioning. With acoustic sectioning, depth information was derived by converting the time-resolved PA signals into depth-resolved signals based on the known speed of sound of the acoustic coupling medium (1485 m/s), while optical sectioning was achieved by focusing and raster-scanning of the excitation laser at different distances from the distal tip of the MMF. Phantoms comprising carbon fibres were used for 3D imaging with a raster-scanning step of 1 $\mu$m. The gap between each two imaging planes was 5 $\mu$m in optical sectioning mode. The intensity profiles of the carbon fibre images along the depth direction were used to calculate the axial resolutions of 3D-PA imaging with both the acoustic sectioning and optical sectioning modes.

### 2.4. Photoacoustic imaging of red blood cells

To evaluate the potential of the needle probe for imaging biological samples, PA imaging was performed on a mouse blood smear on a coverslip. Mouse blood was obtained from culled mice. The procedures involving mice were ethically reviewed and carried out in accordance with the Animals (Scientific Procedures) Act 1986 (ASPA) UK Home Office regulations governing animal experimentation. After MMF characterisation, the sample was placed in front of the MMF tip with a distance of ∼ 10 $\mu$m. Laser was scanned at 5 focal planes with an interval of 5 $\mu$m, covering a total depth of 20 $\mu$m.

Video mosaicing was employed to enlarge the FOV of the probe. With mosaicing imaging, MIP PA images were acquired in real-time during the horizontal translation of the needle probe, whilst adjacent images were registered and stitched according to the their common features at the overlapped margins. A fast registration algorithm that performs optimised cross-correlations using discrete Fourier transformations [30] was used to calculate the displacement between two sequential frames for frame registration. The calculation was based on the open-source MATLAB code provided by the authors of Ref. [30]. The dimensions of the obtained PA images were upsampled by 10 times with bicubic interpolation to improve the precision of the image stitching. For creating the real-time mosaicing image, a large zero-value background was generated and the image intensity values of the first frame was added to the region corresponding to the initial needle location. After acquisition of the subsequent frame, the displacement between the current frame and the last frame was calculated via the fast registration algorithm, and the current frame was superimposed into the same background at an offset determined by the displacement. This registration and stitching process repeated until the entire area corresponding to the background was completely scanned.

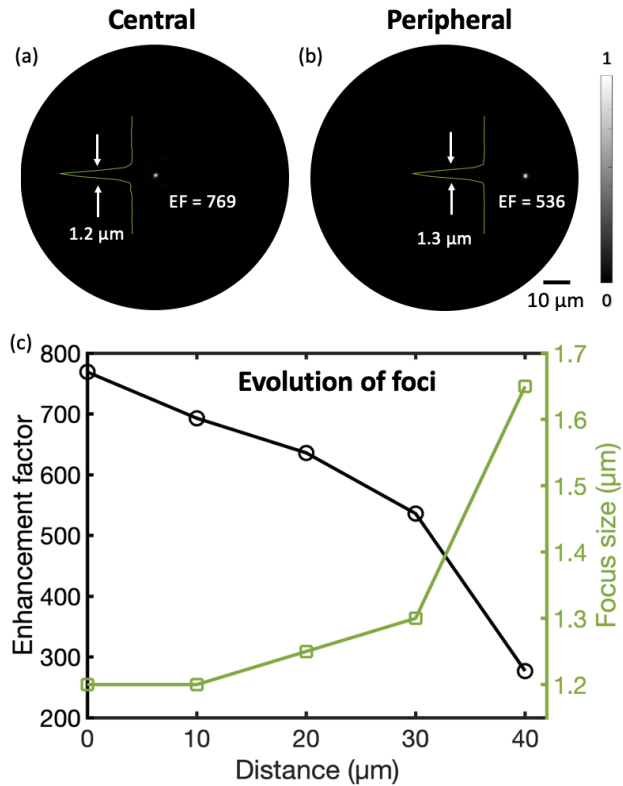

Fig. 2. Focusing performance through a multimode fibre. (a) An example of foci at central region of the fibre tip. Inset is the intensity profile across the centre of the focus indicating that the diameter of the focus is 1.2 $\mu$m. (b) An example of foci at peripheral region of the fibre tip. Inset is the intensity profile across the centre of the focus indicating that the diameter of the focus is 1.3 $\mu$m. (c) The evolution of enhancement factor (EF) and size of laser foci with varying radial distances to the centre of the fibre tip.

## 3. Results

### 3.1. Spatial resolution

The light focus was characterised as the point spread function of the endomicroscopy imaging system. Two examples of light foci at the central and peripheral positions are shown in Fig. 2 (a) and (b), respectively. The focus at the central region had an EF of 769 (~8.9% energy in the focal area) and a diameter of 1.2 $\mu$m. The EF decreased and size of the focus increased with the focus moved away from the centre: when the focus was 30 $\mu$m away from the centre, the EF declined to 536 (~7.7%) and size increased to 1.3 $\mu$m (Fig 2c). In addition, the focusing performance at the very edge of the MMF (50 $\mu$m away from the centre) showed a significantly degradation and thus it is not presented in Fig. 2 (c).

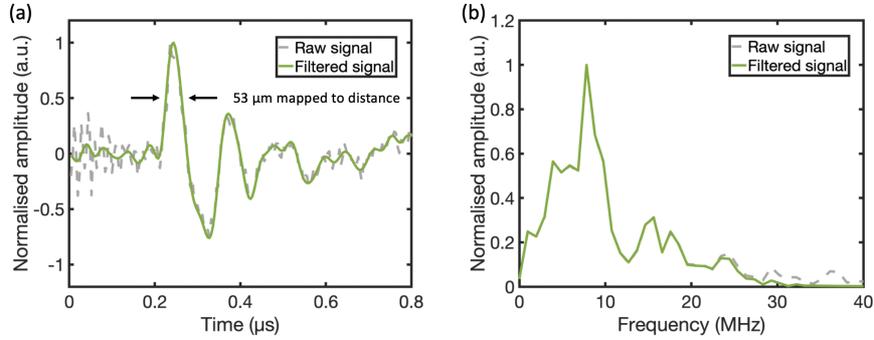

Fig. 3. Characterisation of a received photoacoustic signal. (a) A representative photoacoustic signal generated from a carbon fibre and acquired by the fibre-optic microresonator ultrasound sensor (before and after frequency filtering). The full width at half maximum of the signal was mapped to a distance of 53 $\mu$m. (b) Frequency spectra of the photoacoustic signal before and after frequency filtering.

The waveform of a PA signal generated from a carbon fibre and received by the fibre-optic microresonator ultrasound sensor is shown in Fig. 3 (a). The signal was denoised with a low-pass filter with a cut-off frequency of 20 MHz. The FWHM of the positive peak was mapped into a distance of 53 $\mu$m as an estimation of the axial resolution with acoustic sectioning. The frequency spectrum of the PA signal is shown in Fig. 3 (b). The PA signal had a central frequency of 8 MHz, and a -6 dB bandwidth of 6.5 MHz.

The results of the lateral resolution measurement are shown in Fig 4. MIP images of a resolution target in a $\phi$100 $\mu$m area were achieved with a scanning step of 0.5 $\mu$m (Fig. 4a). The lateral resolutions were estimated as 1.2 $\mu$m and 1.25 $\mu$m at central and peripheral regions, respectively (Fig. 2b and c), which were consistent with the sizes of the optical foci through the MMF. The DMD was operated at 22.7 kHz and the image comprised 31500 pixels. Thus, it took ~1.4 s for the acquisition of such an image. The same sample was also imaged with a scanning step of 1 $\mu$m to reduce the total number of scan positions for a higher imaging speed. As shown in Fig. 4 (d-f), the image quality slightly degraded, with the lateral resolution at central and peripheral regions declined to 1.25 $\mu$m and 1.4 $\mu$m, respectively, whilst the acquisition time of such an image was reduced to ~0.35 s.

The results of the axial resolution measurement are shown in Fig. 5. Volumetric rendering for 3D-PA images of carbon fibre phantoms (acoustic sectioning and optical sectioning) was achieved with the isosurface function in MATLAB as shown in Fig. 5 (a) and (e), respectively. 3D rendering at different views are shown in Video 1 and Video 2. With acoustic sectioning, a MIP image is shown in Fig. 5 (b), and the YZ cross-section plane along the dash green line across two carbon fibres is shown in Fig. 5 (c). The range of depth was estimated to be 266 $\mu$m with the speed of sound (1485 m/s). The ESF profile of a carbon fibre along the red line in Fig. 5 (c) was fitted with a Gaussian function and the FWHM of the ESF was measured to be 50 $\mu$m, which is consistent with the mapped FWHM of the positive peak of the ultrasound signal (Fig. 3a). 3D imaging was also achieved with optical sectioning by scanning the laser focus at 21 planes with an interval of 5 $\mu$m. The XY plane of a carbon fibre phantom at one depth and the YZ sectioning plane along the dash green line are shown in Fig. 5 (f) and (g), respectively. The FWHM of the intensity profile was measured to be 25 $\mu$m, which is consistent with the estimated depth of field (22 $\mu$m) of the used fibre. The raster-scan step was set to be 1 $\mu$m and the DMD was operated at a rate of 22.7 kHz. As such, the imaging speed of 3D imaging with acoustic sectioning was ~0.35 s for per image. Since the optical sectioning method comprised 21 focal planes covering a depth range of 100 $\mu$m, it took 7.3 s for the acquisition of a 3D image with the optical sectioning.

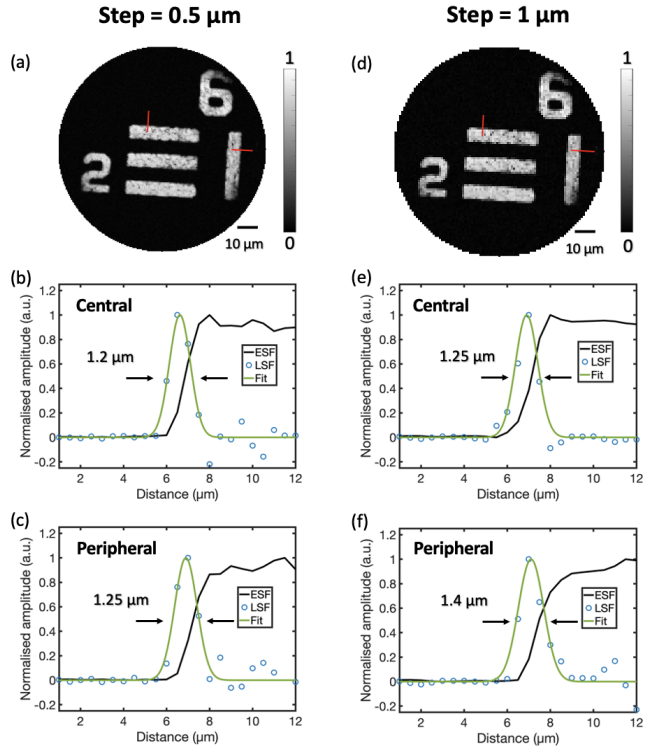

Fig. 4. Characterisation of lateral resolution. (a) Photoacoustic maximum-intensity-projection image of a resolution target with a scanning step of 0.5 $\mu$m. (b) and (c) are edge spread functions (ESF) and line spread functions (LSF) obtained along the red lines in (a) at central and peripheral regions, respectively. (d) Photoacoustic maximum-intensity-projection image of the same sample with a scanning step of 1 $\mu$m. (e) and (f) are ESF and LSF obtained along the red lines in (d) at central and peripheral regions, respectively. Note that 10 adjacent lines from (a) and 5 ones from (d) were summed to suppress noise, while only one of the profiles taken across the edge of the resolution target is shown.

### 3.2. Photoacoustic imaging of red blood cells

3D imaging with optical sectioning was demonstrated by imaging a red blood cell (RBC) smear sample (Fig. 6a). As shown in Fig. 6 (b), a stack of MIP images were achieved by raster-scanning the laser focus at 5 different depths in front of the MMF tip via wavefront shaping. The interval between two adjacent depths was set as 5 $\mu$m. An example of a single RBC (in the red box in Fig. 6a) is shown as a volumetric rendering in Fig. (c), which clearly visualised the biconcave structure of the cell. Images from more views are shown in Video 3. At the centre of the MMF tip, the total energy at the optical focus (1.2 $\mu$m in diameter) was measured as ∼20 nJ which was 8.9% of the total output of the MMF, leading to an optical fluence of 1.7 $J/cm^2$. Each image in Fig. 6 (b) had a diameter of 100 $\mu$m and comprised ∼31500 pixels with a raster-scan step of 0.5 $\mu$m. With the DMD operating at 22.7 kHz, the time required for single MIP image acquisition was ∼1.4 s and thus, it took ∼7 s for acquiring the whole 3D volume with optical sectioning covering a depth of 20 $\mu$m.

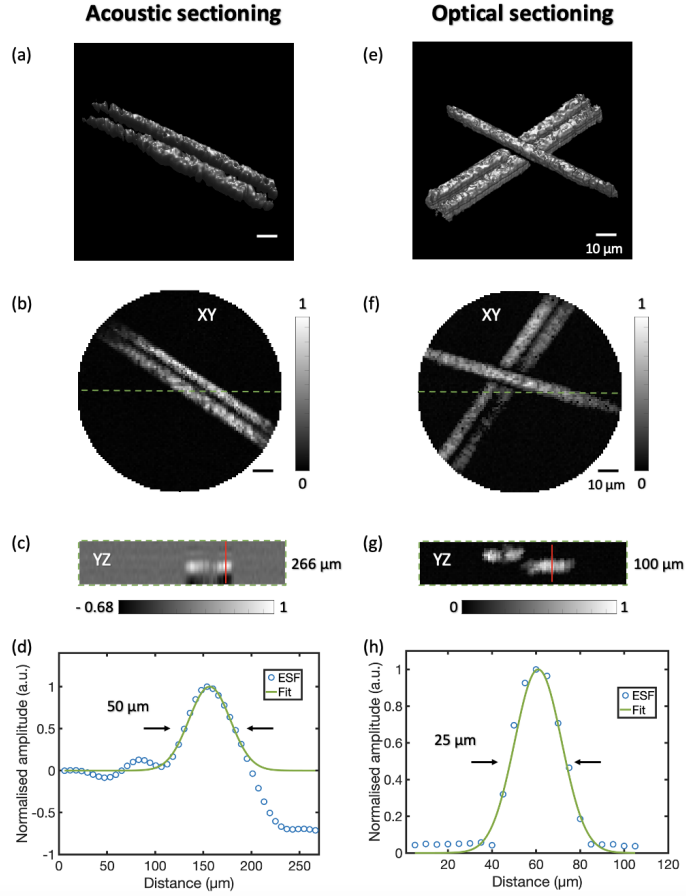

Fig. 5. Characterisation of axial resolution. (a) Volumatric rendering of a carbon fibre phantom with acoustic sectioning (Visualisation from multiple views is shown in Video 1). (b) Photoacoustic maximum-intensity-projection image of the phantom with acoustic sectioning. (c) The cross-section plane along the dash green line in (b). (d) The edge spread function across the red line in (c). (e) Volumatric rendering of a carbon fibre phantom with optical sectioning (Visualisation from multiple views is shown in Video 2). (f) Photoacoustic maximum-intensity-projection image of the phantom at one focal plane. (g) The cross-section plane along the dash green line in (f). (h) The edge spread function across the red line in (g). D, diameter; ESP, edge spread function.

Mosaicing was implemented by translating the endomicroscopy probe over a mouse blood smear sample and stitching frames in real-time. An example of real-time mosaicing is shown in Video 4, and the last frame of the mosaicing video is shown in Fig. 7 with the needle translation shown in the inset. A single frame of OR-PAM imaging covered a 100 $\mu$m-in-diameter area, the step of the raster-scan was 1 $\mu$m, and each frame comprised 7850 pixels. So, the speed for single frame acquisition was ~3 frames per second with the DMD operating at 22.7 kHz. The mosaicing image shown in Fig. 7 was obtained with 40 consecutive OR-PAM images and covered an area of around 100 $\mu$m × 250 $\mu$m. The consecutive frames were superimposed onto the large background and as a result, the image in the overlapped regions was improved owing to the averaging effect. The biconcave structures of RBCs were clearly visualised. Further, as image quality did not seem to have been substantially affected during the translation of the needle probe, the imaging probe has demonstrated a high degree of resistance to modest fibre bending.

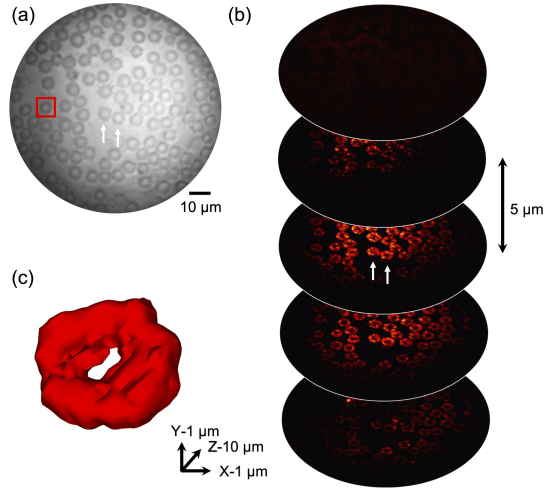

Fig. 6. Three-dimension photoacoustic endomicroscopy imaging of *ex vivo* mouse blood cells with optical sectioning. (a) Optical microscopy of a mouse blood smear sample. (b) Slices through the optical sectioning planes within a range of 20 $\mu$m with an interval of 5 $\mu$m. (c) An example of volumetric rendering of a single red blood cell. Visualisation from different views of the same cell is shown in Video 3.

## 4. Discussion

In this work, we developed a highly miniaturised, high-speed, forward-reviewing, PA endomicroscopy probe integrated within the cannula of a 20 gauge medical needle for guiding minimally invasive procedures. This probe is based on light focusing through a MMF assisted by high-speed wavefront shaping using a DMD and a fibre-optic planar-concave microresonantor ultrasound sensor for signal detection. The fibre-optic ultrasound sensor is well suited to the development of such an ultrathin probe as this miniature sensor provides a high acoustic sensitivity, broad frequency bandwidth, and a near omni-directional response. In comparison, to match the fibre sensor sensitivity, a piezoelectrical transducer would have to be made so large that it would be highly directional, which represents fundamental limitations for miniaturisation.

The high-speed imaging capability was enabled by using a high-speed DMD for wavefront shaping. The MMF was characterised with a RVITM-based approach [28]. It took around 3 min for characterisation at one focal plane and thus fibre characterisation for optical sectioning at 21 planes took around 1 h. With the DMD operating at 22.7 kHz for imaging of RBCs, it improved the image acquisition speed by more than 2 orders of magnitude compared to that achieved with a LC-SLM (60 Hz) [19]. Apart from imaging speed, the high scanning speed also allowed a denser spatial sampling (more pixels in the same area), and hence enabled a sub-cellular spatial resolution with comparable fidelity to benchtop PAM systems [31]. Deep learning could also be used to improve the imaging speed by increasing the scanning step size and hence reducing the total number of scans without sacrificing the spatial resolution as demonstrated in previous studies [32–34].

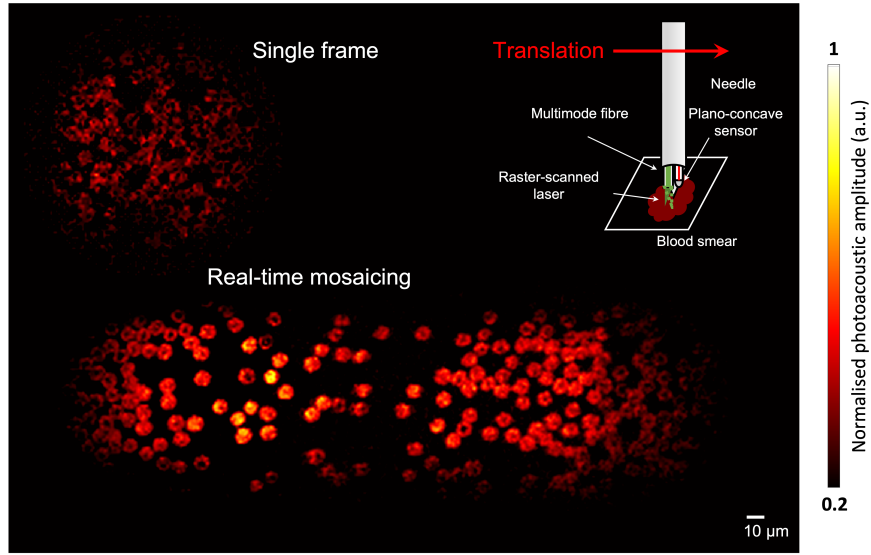

Fig. 7. Mosaicing imaging of a mouse blood smear sample over an area of 100 $\mu$m × 250 $\mu$m. Each single frame covers an area with a diametre of 100 $\mu$m with a raster-scanning step size of 1 $\mu$m.

Volumetric PA imaging was realised by both acoustic sectioning and optical sectioning. With acoustic sectioning, the axial resolution depends on the frequency bandwidth of the ultrasound sensor. For example, with the current ultrasound sensor, the axial resolution was measured as around 50 $\mu$m, which is worse than that enabled by optical sectioning (25 $\mu$m). However, the imaging acquisition time increases with the number of the focal planes for optical sectioning, which is unfavourable for *in vivo* applications. Recent studies have shown that further increasing the frequency bandwidth of the ultrasound sensor can achieve even higher axial resolution than that with optical sectioning [31, 35, 36].

Although the optical fluence at the light foci (1.7 $J/cm^2$) is comparable to those commonly used in OR-PAM [19, 35, 37–39], it is much higher than the maximum permissible exposure (MPE) at 532 nm (20 $mJ/cm^2$) according to the ANSI standard [40]. Advanced denoising algorithms such as recent developments in deep learning could be used to bring down the required optical fluence at the foci [41]. However, the MPE is defined for skin but not inside the body. In the absence of regulatory guidance it is likely that practical use in endoscopy will require precursor safety studies (e.g. examining tissue histology for damage following different exposure levels) for specific tissues or organs on a case-by-case basis [42, 43].

Focusing light though a MMF by wavefront shaping usually requires a stationary fibre during imaging. However, as reported by Flaes et al. in 2018 [44], graded-index MMF showed a high robustness to fibre bending deformation, which allows raster-scan-based imaging even when the fibre was bent with large curvatures. A 140 $\mu$m, 0.29 NA graded-index fibre was employed in our system, and its robustness to fibre bending has been reported in our previous study [23]. Here with mosaicing imaging, although the needle probe was translated, the imaging performance was not substantially degraded, suggesting that this graded-index fibre was resistant to modest fibre bending. However, a systematical study of the robustness of this kind of fibre to complex shape changes is needed.

This work focused on the development of PA endomicroscopy, however, the developed high-speed wavefront shaping technology can be used for other endoscopic imaging modalities such as fluorescence microscopy [22, 45, 46], Raman microscopy [47, 48], and two-photon microscopy [49]. In recent studies [45, 46], fluorescence microscopy through MMFs has been demonstrated with an *in vivo* animal model, which is promising to provide complementary contrast to PA imaging. These imaging modalities could also be combined with PA endomicroscopy by sharing the same MMF for light delivery to provide complementary information of tissue. In the future, multispectral excitation could be implemented to provide functional information of tissue such as blood oxygen saturation.

## 5. Conclusions

In summary, we developed a highly miniaturised, high-speed, forward-reviewing, optical-resolution PA endomicroscopy probe based on a MMF and a highly sensitive fibre-optic microresonator ultrasound sensor. High-fidelity 3D images of mouse red blood cells were acquired at an unprecedented speed. This needle probe thus holds the potential for providing 3D micro-structural, functional and molecular information of tissue at sub-celluar spatial resolution *in situ* for guiding minimally invasive procedures such as tumour biopsy.

**Funding.** This project was funded in whole, or in part, by the Academy of Medical Sciences/the Wellcome Trust/ the Government Department of Business, Energy and Industrial Strategy/the British Heart Foundation/Diabetes UK Springboard Award [SBF006/1136], Wellcome Trust (203148/Z/16/Z, WT101957), Engineering and Physical Sciences Research Council (NS/A000027/1, NS/A000049/1), and ERC Advanced Grant 74119.

**Acknowledgments.** The authors thank Dr. Sunish Mathews, Dr. Fang-Yu Lin and Dr. Miguel Xochicale for their helps with setting up the fibre-optic sensor console.

**Disclosures.** The authors declare that they have no known competing financial interests or personal relationships. T.V. is co-founder and shareholder of Hypervision Surgical Ltd, London, UK. He is also a shareholder of Mauna Kea Technologies, Paris, France. P.B. and E.Z. are shareholders in DeepColor Imaging SAS.

**Data Availability Statement.** Code and data generated during this study are available from the corresponding author on reasonable request.

## 6. References

**References**

1. T. D. Wang and J. Van Dam, "Optical biopsy: a new frontier in endoscopic detection and diagnosis," Clin. gastroenterology hepatology **2**, 744–753 (2004).
2. M. J. Gora, M. J. Suter, G. J. Tearney, and X. Li, "Endoscopic optical coherence tomography: technologies and clinical applications," Biomed. optics express **8**, 2405–2444 (2017).
3. J. Mavadia, J. Xi, Y. Chen, and X. Li, "An all-fiber-optic endoscopy platform for simultaneous oct and fluorescence imaging," Biomed. optics express **3**, 2851–2859 (2012).
4. P. Beard, "Biomedical photoacoustic imaging," Interface focus **1**, 602–631 (2011).
5. V. Ntziachristos and D. Razansky, "Molecular imaging by means of multispectral optoacoustic tomography (msot)," Chem. reviews **110**, 2783–2794 (2010).
6. M. Xu and L. V. Wang, "Photoacoustic imaging in biomedicine," Rev. scientific instruments **77**, 041101 (2006).
7. T. Zhao, A. E. Desjardins, S. Ourselin, T. Vercauteren, and W. Xia, "Minimally invasive photoacoustic imaging: Current status and future perspectives," Photoacoustics **16**, 100146 (2019).

8. J. Zhou and J. V. Jokerst, "Photoacoustic imaging with fiber optic technology: A review," Photoacoustics p. 100211 (2020).
9. K. Jansen, A. F. Van Der Steen, H. M. van Beusekom, J. W. Oosterhuis, and G. van Soest, "Intravascular photoacoustic imaging of human coronary atherosclerosis," Opt. letters **36**, 597–599 (2011).
10. J.-M. Yang, C. Favazza, R. Chen, J. Yao, X. Cai, K. Maslov, Q. Zhou, K. K. Shung, and L. V. Wang, "Simultaneous functional photoacoustic and ultrasonic endoscopy of internal organs in vivo," Nat. medicine **18**, 1297–1302 (2012).
11. P. Wang, T. Ma, M. N. Slipchenko, S. Liang, J. Hui, K. K. Shung, S. Roy, M. Sturek, Q. Zhou, Z. Chen *et al.*, "High-speed intravascular photoacoustic imaging of lipid-laden atherosclerotic plaque enabled by a 2-khz barium nitrite raman laser," Sci. reports **4**, 1–7 (2014).
12. R. Ansari, E. Z. Zhang, A. E. Desjardins, and P. C. Beard, "All-optical forward-viewing photoacoustic probe for high-resolution 3d endoscopy," Light. Sci. & Appl. **7**, 1–9 (2018).
13. R. Ansari, E. Z. Zhang, A. E. Desjardins, and P. C. Beard, "Miniature all-optical flexible forward-viewing photoacoustic endoscopy probe for surgical guidance," Opt. Lett. **45**, 6238–6241 (2020).
14. P. Hajireza, W. Shi, and R. Zemp, "Label-free in vivo fiber-based optical-resolution photoacoustic microscopy," Opt. letters **36**, 4107–4109 (2011).
15. P. Shao, W. Shi, P. H. Reza, and R. J. Zemp, "Integrated micro-endoscopy system for simultaneous fluorescence and optical-resolution photoacoustic imaging," J. Biomed. Opt. **17**, 076024 (2012).
16. G. Li, Z. Guo, and S.-L. Chen, "Miniature probe for forward-view wide-field optical-resolution photoacoustic endoscopy," IEEE Sensors J. **19**, 909–916 (2018).
17. I. N. Papadopoulos, O. Simandoux, S. Farahi, J. Pierre Huignard, E. Bossy, D. Psaltis, and C. Moser, "Optical-resolution photoacoustic microscopy by use of a multimode fiber," Appl. Phys. Lett. **102**, 211106 (2013).
18. N. Stasio, A. Shibukawa, I. N. Papadopoulos, S. Farahi, O. Simandoux, J.-P. Huignard, E. Bossy, C. Moser, and D. Psaltis, "Towards new applications using capillary waveguides," Biomed. optics express **6**, 4619–4631 (2015).
19. S. Mezil, A. M. Caravaca-Aguirre, E. Z. Zhang, P. Moreau, I. Wang, P. C. Beard, and E. Bossy, "Single-shot hybrid photoacoustic-fluorescent microendoscopy through a multimode fiber with wavefront shaping," Biomed. Opt. Express **11**, 5717–5727 (2020).
20. D. Wang, E. H. Zhou, J. Brake, H. Ruan, M. Jang, and C. Yang, "Focusing through dynamic tissue with millisecond digital optical phase conjugation," Optica **2**, 728–735 (2015).
21. S. Turtaev, I. T. Leite, K. J. Mitchell, M. J. Padgett, D. B. Phillips, and T. Čižmár, "Comparison of nematic liquid-crystal and dmd based spatial light modulation in complex photonics," Opt. express **25**, 29874–29884 (2017).
22. A. M. Caravaca-Aguirre and R. Piestun, "Single multimode fiber endoscope," Opt. express **25**, 1656–1665 (2017).
23. T. Zhao, M. T. Ma, S. Ourselin, T. Vercauteren, and W. Xia, "Video-rate dual-modal photoacoustic and fluorescence imaging through a multimode fibre towards forward-viewing endomicroscopy," Photoacoustics **25**, 100323 (2022).
24. E. Z. Zhang and P. C. Beard, "A miniature all-optical photoacoustic imaging probe," in *Photons plus ultrasound: imaging and sensing 2011,* vol. 7899 (SPIE, 2011), pp. 291–296.
25. J. A. Guggenheim, J. Li, T. J. Allen, R. J. Colchester, S. Noimark, O. Ogunlade, I. P. Parkin, I. Papakonstantinou, A. E. Desjardins, E. Z. Zhang *et al.*, "Ultrasensitive plano-concave optical microresonators for ultrasound sensing," Nat. Photonics **11**, 714–719 (2017).
26. T. Čižmár and K. Dholakia, "Shaping the light transmission through a multimode optical fibre: complex transformation analysis and applications in biophotonics," Opt. Express **19**, 18871–18884 (2011).
27. T. Zhao, S. Ourselin, T. Vercauteren, and W. Xia, "High-speed photoacoustic-guided wavefront shaping for focusing light in scattering media," Opt. Lett. **46**, 1165–1168 (2021).
28. T. Zhao, S. Ourselin, T. Vercauteren, and W. Xia, "Focusing light through multimode fibres using a digital micromirror device: a comparison study of non-holographic approaches," Opt. Express **29**, 14269–14281 (2021).
29. A. Descloux, L. V. Amitonova, and P. W. Pinkse, "Aberrations of the point spread function of a multimode fiber due to partial mode excitation," Opt. Express **24**, 18501–18512 (2016).
30. M. Guizar-Sicairos, S. T. Thurman, and J. R. Fienup, "Efficient subpixel image registration algorithms," Opt. letters **33**, 156–158 (2008).
31. B. Dong, H. Li, Z. Zhang, K. Zhang, S. Chen, C. Sun, and H. F. Zhang, "Isometric multimodal photoacoustic microscopy based on optically transparent micro-ring ultrasonic detection," Optica **2**, 169–176 (2015).
32. T. Zhao, M. Shi, S. Ourselin, T. Vercauteren, and W. Xia, "Ai-enabled high-speed photoacoustic endomicroscopy through a multimode fibre," in *SPIE Proceedings Volume 11960, Photons Plus Ultrasound: Imaging and Sensing 2022;*, (2022), p. 119600L.
33. T. Vu, A. DiSpirito III, D. Li, Z. Wang, X. Zhu, M. Chen, L. Jiang, D. Zhang, J. Luo, Y. S. Zhang *et al.*, "Deep image prior for undersampling high-speed photoacoustic microscopy," Photoacoustics **22**, 100266 (2021).
34. A. DiSpirito, D. Li, T. Vu, M. Chen, D. Zhang, J. Luo, R. Horstmeyer, and J. Yao, "Reconstructing undersampled photoacoustic microscopy images using deep learning," IEEE transactions on medical imaging **40**, 562–570 (2020).
35. T. J. Allen, O. Ogunlade, E. Zhang, and P. C. Beard, "Large area laser scanning optical resolution photoacoustic microscopy using a fibre optic sensor," Biomed. optics express **9**, 650–660 (2018).
36. R. Shnaiderman, G. Wissmeyer, O. Ülgen, Q. Mustafa, A. Chmyrov, and V. Ntziachristos, "A submicrometre silicon-on-insulator resonator for ultrasound detection," Nature **585**, 372–378 (2020).
37. S. Hu, K. Maslov, and L. V. Wang, "Second-generation optical-resolution photoacoustic microscopy with improved sensitivity and speed," Opt. letters **36**, 1134–1136 (2011).

38. J. Yao, L. Wang, J.-M. Yang, L. S. Gao, K. I. Maslov, L. V. Wang, C.-H. Huang, and J. Zou, "Wide-field fast-scanning photoacoustic microscopy based on a water-immersible mems scanning mirror," J. biomedical optics **17**, 080505 (2012).
39. G. Wissmeyer, D. Soliman, R. Shnaiderman, A. Rosenthal, and V. Ntziachristos, "All-optical optoacoustic microscope based on wideband pulse interferometry," Opt. letters **41**, 1953–1956 (2016).
40. S. L. I. of America, "Ansi z136.1 – 2014 american national standard for safe use of lasers," Am. Natl. Standards Institute, Inc. (2013).
41. M. J. Bianco, P. Gerstoft, J. Traer, E. Ozanich, M. A. Roch, S. Gannot, and C.-A. Deledalle, "Machine learning in acoustics: Theory and applications," The J. Acoust. Soc. Am. **146**, 3590–3628 (2019).
42. J. Huang, A. Wiacek, K. M. Kempski, T. Palmer, J. Izzi, S. Beck, and M. A. L. Bell, "Empirical assessment of laser safety for photoacoustic-guided liver surgeries," Biomed. Opt. Express **12**, 1205–1216 (2021).
43. T. Sowers, D. VanderLaan, A. Karpiouk, D. Onohara, S. Schmarkey, S. Rousselle, M. Padala, and S. Emelianov, "In vivo safety study using radiation at wavelengths and dosages relevant to intravascular imaging," J. biomedical optics **27**, 016003 (2022).
44. D. E. B. Flaes, J. Stopka, S. Turtaev, J. F. De Boer, T. Tyc, and T. Čižmár, "Robustness of light-transport processes to bending deformations in graded-index multimode waveguides," Phys. review letters **120**, 233901 (2018).
45. S. Turtaev, I. T. Leite, T. Altwegg-Boussac, J. M. Pakan, N. L. Rochefort, and T. Čižmár, "High-fidelity multimode fibre-based endoscopy for deep brain in vivo imaging," Light. Sci. & Appl. **7**, 1–8 (2018).
46. S. Ohayon, A. Caravaca-Aguirre, R. Piestun, and J. J. DiCarlo, "Minimally invasive multimode optical fiber microendoscope for deep brain fluorescence imaging," Biomed. optics express **9**, 1492–1509 (2018).
47. J. Trägårdh, T. Pikálek, M. Šerỳ, T. Meyer, J. Popp, and T. Čižmár, "Label-free cars microscopy through a multimode fiber endoscope," Opt. express **27**, 30055–30066 (2019).
48. I. Gusachenko, M. Chen, and K. Dholakia, "Raman imaging through a single multimode fibre," Opt. Express **25**, 13782–13798 (2017).
49. E. E. Morales-Delgado, D. Psaltis, and C. Moser, "Two-photon imaging through a multimode fiber," Opt. express **23**, 32158–32170 (2015).